\documentclass[
fleqn]{elsart}
\usepackage{amsmath,amssymb,dcolumn}

\def\be{\begin{equation}}
\def\ee{\end{equation}}
\def\bea{\begin{eqnarray}}
\def\eea{\end{eqnarray}}
\def\nn{\nonumber}
\def\J{\mathcal{J}}

\begin{document}
\begin{frontmatter}
\begin{flushright}
hep-th/0701002 v2
\end{flushright}

\title{Hawking radiation from rotating black holes in anti-de Sitter spaces via gauge and gravitational
anomalies}
\author[1]{Qing-Quan Jiang},
\ead{jiangqingqua@126.com}
\author[2]{Shuang-Qing Wu}
\ead{sqwu@phy.ccnu.edu.cn}
\address[1]{Institute of Particle Physics, Central China Normal University, Wuhan, Hubei 430079,
People's Republic of China}
\address[2]{College of Physical Science and Technology, Central China Normal University, Wuhan,
Hubei 430079, People's Republic of China}

\begin{abstract}
Robinson-Wilczek's recent work, which treats Hawking radiation as a compensating flux to cancel
gravitational anomaly at the horizon of a Schwarzschild-type black hole, is extended to study
Hawking radiation of rotating black holes in anti-de Sitter spaces, especially that in dragging
coordinate system, via gauge and gravitational anomalies. The results show that
in order to restore gauge invariance and general coordinate covariance at the quantum level in
the effective field theory, the charge and energy flux by requiring to cancel gauge and gravitational
anomalies at the horizon, must have a form equivalent to that of a $(1+1)$-dimensional blackbody
radiation at Hawking temperature with an appropriate chemical potential.
\end{abstract}

\end{frontmatter}

\newpage

\section{Introduction}\label{intro}

The study of Hawking radiation continued to attract a great deal of attention in theoretical physics
since it was first proposed in 1974 \cite{SWH}. There exist several derivations of Hawking radiation
in the literature (see, for example, \cite{PW,JWC,GH,CF}). Among them, one method to derive Hawking
radiation is to calculate the energy momentum tensor flux in the black hole background. The
Christensen-Fulling's seminal work \cite{CF} showed that the trace anomaly, an anomaly in conformal
symmetry, can derive important constraints on the energy momentum tensor of quantum fields in a black
hole background. Due to the anomaly in conformal symmetry, the flux of Hawking radiation can be treated
as that of the trace of the energy momentum tensor. However, their observation was based upon several
assumptions: firstly, the background was limited to $(1+1)$ dimensions; secondly, the fields were
massless; and finally, there was no back-scattering effect for the massless particles in $(1+1)$
dimensions. Therefore, Hawking radiation appears as a rather special phenomenon in this approach.
Recently, Robinson and Wilczek \cite{RW} proposed another novel approach which ties Hawking radiation
to the cancellation of gravitational anomaly at the horizon of the Schwarzschild-type black hole.
Specifically, the scalar field theory in an arbitrary dimensional black hole spacetimes can be reduced
to that in a $(1+1)$-dimensional spacetime by using a dimensional reduction technique. In the effective
two-dimensional reduction, each partial wave of the scalar field exhibits a general coordinate
symmetry. However, when omitting the classically irrelevant ingoing modes at the horizon, the
effective theory becomes chiral, and the anomaly with respect to general coordinate symmetry
arises to impose great constraints on the energy momentum tensor\footnote{In fact, the effective
field theory is only basing on the observable physics and defined outside the horizon
of the black hole. An observer who lives outside a Schwarzschild-type black hole with finite
energy can not observe the physics beyond the horizon of the black hole since the horizon is a
null-hypersurface. As the global Killing vector that describes the symmetry of the spacetime is
only time-like outside its horizon, we can define the energy of quantum states in an effective theory
that only describes observable physics. However, the energy momentum tensor calculated by this definition
is divergent at the horizon due to a pile up of the horizon-skimming modes. Thus, the effective theory can
properly describe the observable physics with these modes integrated out. Although such formulated effective
field theory no longer has observable divergence, it contains gravitational anomaly that imposes great constraints
on the energy momentum tensor.}. Meanwhile, they pointed out that if demanding general coordinate covariance at the
quantum level to hold in the effective field theory, gravitational anomaly can be cancelled by the
quantum effect of the classically irrelevant ingoing mode, and the flux of energy momentum tensor
is equal to that of (1+1)-dimensional blackbody radiation at Hawking temperature.

In the case of the charged or rotating black holes in arbitrary dimensions, the quantum field near the black
hole horizon can be interpreted, by a dimensional reduction technique, as that in the backgrounds of the
dilaton (whose contribution to Hawking flux is dropped due to the static background), $(1+1)$-dimensional
metric, and the gauge field. However, the effective gauge potential corresponding to the gauge field in the
two-dimensional reduction for the charged black holes is originated from the electric field of the black
holes, while for rotating black holes it is due to the $U(1)$ gauge symmetry which is associated with
the axisymmetry of the black hole. After neglecting the classically irrelevant ingoing modes near the
horizon, the effective two-dimensional theory becomes chiral there, and contains gauge and gravitational
anomalies to give constraints on the gauge current and the energy momentum tensor. Hawking flux from
these two types of black holes can be determined in terms of the values of anomalies at the horizon
by demanding gauge invariance and general coordinate covariance at the quantum level \cite{IUW1,MS,IUW2,VD}.
Obviously, the derivation of Hawking radiation in this way can be more universal since it is only dependent
on the conditions of anomaly cancellation at the horizon, no matter what the black holes in arbitrary
dimensions are rotating or charged.

On the other hand, properties of black holes in anti-de Sitter (AdS) spaces especially those of thermodynamics
\cite{BHT} have been investigated thoroughly in recent years within the context of the AdS/CFT correspondence
\cite{AdSCFT}. Thus it is interesting to investigate the Hawking radiation of rotating black holes in AdS
spaces from the viewpoint of anomaly cancellation. In this article, we shall
study Hawking radiation from rotating black holes in AdS spaces via gauge and gravitational anomalies
at the horizon. By using a dimensional reduction technique, we can reduce the scalar field near the horizon
of the black holes to that in the backgrounds of the dilaton, $(1+1)$-dimensional metric, and the gauge field.
Here the contribution of dilaton background to the flux of Hawking radiation is dropped due to a static
background. It should be noted that although the dilaton and the conformal factor of the metric share a
common form, the gauge potential in the two-dimensional reduction of a Kerr-AdS black hole is originated
from the induced symmetry associated with the isometry along the $\phi$-direction, and the charge of the
field is given by the azimuthal quantum number $m$, while that of a Kerr-Newman-AdS black hole is composed
of one primary gauge potential originated from the electric field of the black hole and another $U(1)$
gauge potential associated with the axisymmetry of the black hole. Since the matter field in the ergosphere
near the horizon must be dragged by the gravitational field of a spinning source because there exists a
frame-dragging effect of the coordinate in the rotating spacetime, so it is also physically reasonable to
investigate this issue in the dragging coordinate system. When do so, one can observe that the $U(1)$ gauge
symmetry with respect to the isometry along this $\phi$-direction may not be incorporated in the gauge symmetry
of the two-dimensional reduction of the Kerr-Newman-AdS black holes. In all, after omitting the classically
irrelevant ingoing modes at the horizon, the effective chiral theory for a Kerr-AdS black hole contains a
gravitational anomaly and a $U(1)$ gauge anomaly, and that for a Kerr-Newman-AdS black hole is composed of
a gravitational anomaly and two gauge anomalies. However, in the case of a Kerr-Newman-AdS black hole in the
dragging coordinate system, the effective two-dimensional chiral theory does not contain the $U(1)$ gauge
anomaly associated with the induced symmetry originated from the isometry along the $\phi$-direction. An
observer rest at the dragging coordinate system, which behaves like a kind of locally non-rotating coordinate
system, would not observe this $U(1)$ gauge current flux since he is co-rotating with the rotating black hole.
If we restore the Boyer-Lindquist coordinate system from the dragging coordinate system, it is also easy to
calculate the flux of the angular momentum\cite{MS}. In order to maintain gauge invariance or general coordinate
covariance at the quantum level to hold in the effective theory, these anomalies are cancelled by the
fluxes of (1+1) blackbody radiation at Hawking temperature with appropriate chemical potentials.

This paper is outlined as follows. In Sec. \ref{KAdS}, Hawking radiation from a Kerr-AdS black hole is
investigated by gauge and gravitational anomalies at the horizon, we also prove that the fluxes of gauge
current and energy momentum tensor, which are required to cancel the anomalies at the horizon, are exactly
equal to that of a (1+1)-dimensional blackbody radiation at Hawking temperature with an appropriate chemical
potential for an azimuthal angular momentum $m$. In Sec. \ref{KNAdS}, we consider as an example the case of
a Kerr-Newman-AdS black hole the Boyer-Lindquist coordinates and in the dragging coordinates, and prove once
again that if demanding gauge invariance and general coordinate covariance at the quantum level to hold in
the effective theory, the thermal flux of Hawking radiation is capable of cancelling the anomalies at the
horizon. Sec. \ref{dc} ends up with some discussions and conclusions.

\section{Hawking radiation from a Kerr-AdS black hole}\label{KAdS}
The metric of a four-dimensional a Kerr-AdS black hole can be expressed as \cite{CP}
\be
ds^2 = -\frac{\Delta_r}{\rho^2} \Big(dt -\frac{a\sin^2\theta}{\Xi} d\phi\Big)^2
+\frac{\rho^2}{\Delta_r}dr^2 +\frac{\rho^2}{\Delta_\theta}d\theta^2
+\frac{\Delta_\theta \sin^2\theta}{\rho^2}\Big(adt -\frac{r^2 +a^2}{\Xi}d\phi\Big)^2 \, ,
\label{mle}
\ee
where
\bea
\Delta_r &=& (r^2 +a^2)(1 +r^2l^{-2}) -2Mr \, ,
\qquad \Delta_\theta = 1 -a^2l^{-2}\cos^2\theta \, , \nn \\
\rho^2 &=& r^2 +a^2\cos^2\theta \, , \qquad \Xi = 1-a^2l^{-2} \, . \nn
\eea
The outer horizon $(r = r_H)$ is determined by $\Delta_r(r_H) = 0$. Using a dimensional reduction
technique, the scalar field theory in a four-dimensional Kerr-AdS black hole can be reduced to
that in an effective two-dimensional spacetime near the horizon. The action for a scalar field
in the background spacetime (\ref{mle}) is
\bea
S[\varphi] &=& \frac{1}{2}\int d^4x \sqrt{-g}\varphi\nabla^2\varphi  \nn \\
&=& \frac{1}{2}\int dtdrd\theta d\phi \frac{\sin\theta}{\Xi}\varphi\Big[\frac{\Delta_ra^2\sin^2\theta
-\Delta_\theta{(r^2 +a^2)}^2}{\Delta_r\Delta_\theta}\partial_t^2 \nn \\
&& +\frac{\Delta_r -\Delta_\theta(r^2 +a^2)}{\Delta_r\Delta_\theta} 2a\Xi\partial_t\partial_\phi
+\frac{\Delta_r -\Delta_\theta a^2\sin^2\theta}{\Delta_r\Delta_\theta\sin^2\theta}
\Xi^2\partial_\phi^2 \nn \\
&& +\partial_r\big(\Delta_r\partial_r\big) +\frac{1}{\sin\theta}\partial_\theta
\big(\sin\theta\Delta_\theta\partial_\theta\big) \Big]\varphi \, .
\label{sact}
\eea
Performing the partial wave decomposition of the scalar field in the rotating black hole in term of the
spherical harmonics\footnote{Strictly speaking, the angular part of the separated scalar field equation
can be transformed into a form of Heun equation \cite{STU}, but at the horizon it approaches to the
spherical harmonic since the near-horizon geometry has a topology of 2-sphere.} $\varphi = \sum_{l, m}
\varphi_{lm}(t, r)Y_{lm}(\theta, \phi)$ at the horizon, and transforming to the tortoise coordinate,
defined by $dr_*/dr = (r^2+a^2)/\Delta_r \equiv f(r)^{-1}$, the action (\ref{sact}) can be simplified as
\be
S[\varphi] = \frac{r^2 +a^2}{2\Xi}\sum_{l, m}\int dtdr \varphi_{lm}^*
\Big[-\frac{1}{f(r)}{\Big(\partial_t +\frac{i\Xi am}{r^2 +a^2}\Big)}^2
+\partial_r\big(f(r)\partial_r\big)\Big]\varphi_{lm} \, .
\ee
Therefore, after undergoing a dimensional reduction technique near the horizon, each partial wave of
the scalar field $\varphi$ in a four-dimensional Kerr-Ads black hole can be effectively described
by an infinite collection complex scalar field in the background of a $(1+1)$-dimensional metric,
the dilaton $\Psi$, and the $U(1)$ gauge field $A_{\mu}$ as
\bea
g_{tt} &=& -f(r) = -\frac{\Delta_r}{r^2+a^2} \, , \qquad  g_{rr}= \frac{1}{f(r)}  \, , \nn \\
\Psi &=& \frac{r^2 +a^2}{\Xi} \, , \qquad A_t = -\frac{\Xi a}{r^2 +a^2} \, , \quad A_r = 0 \, ,
\eea
where the gauge charge for the $U(1)$ gauge field is the azimuthal quantum number $m$. In the two-dimensional
reduction, when omitting the ingoing modes at the horizon, the effective theory becomes chiral here, but
each partial wave becomes anomalous with respect to gauge and general coordinate symmetries since the
numbers of the ingoing modes and the outgoing modes are no longer identical. To demand gauge invariance
and diffeomorphism covariance to hold in the effective field theory, the fluxes of the $U(1)$ gauge current
and the energy momentum tensor are required to cancel gauge and gravitational anomalies at the horizon,
respectively. In the following discussion, although the two-dimensional effective theory contains a
dilaton background, its contribution to the total flux can be neglected due to the static background.

We first determine the flux of the $U(1)$ gauge current. It is well-known that if the symmetry of action
or the corresponding conservation law, which is valid in the classical theory, is violated in the quantized
version, the anomaly then occurs. In classical theory, to restore the classical action invariance under
gauge transformations, the gauge current must satisfy the conservation equation: $\nabla_{\mu} J^\mu = 0$.
However, in quantum theory, the quantum effect of the ingoing modes could be taken place at the horizon,
a null hypersurface. But if we first omits its effect at the horizon, the gauge current exhibits an
anomaly here, and satisfies the anomalous equation as
\be
\nabla_\mu J^\mu = \frac{m^2}{4\pi\sqrt{-g}}\epsilon^{\mu\nu}\partial_\mu A_\nu \, ,
\ee
where $m$ is the azimuthal quantum number, treated here as the gauge charge of the $U(1)$ gauge field
$A_\nu$. We define the effective field theory outside the horizon of the black hole. In the region $r_H
+\epsilon \leq r$, since there is no anomaly as the fundamental theory does, the $U(1)$ gauge current
satisfies the conservation equation: $\partial_r J_{(o)}^r = 0$, which is obtained from the $(r, \phi)$-component
of the four-dimensional energy momentum tensor (see Eq. (10) in Ref. \cite{IUW2}). However, near the horizon
$r_H \leq r \leq r_H +\epsilon$, after excluding mode's propagation along one light-like direction, the
effective theory is chiral here, but the $U(1)$ gauge current exhibits an anomaly, which means that
$\partial_r J_{(H)}^r = {m^2}\partial_r A_t/{4\pi}$.\footnote{It should be noted that, the $t$-component
of the gauge current is irrelevant for the Hawking flux of the charge.} Under gauge transformations, the
variation of the effective action is $-\delta W = \int dtdr \sqrt{-g}\lambda \nabla_\mu J^\mu$, where
$\lambda$ is a gauge parameter. The consistent current is written as $J^\mu = J_{(o)}^\mu\Theta_+(r)
+J_{(H)}^\mu H(r)$, in which $\Theta_+(r) = \Theta(r -r_H -\epsilon)$ and $H(r) = 1 -\Theta_+(r)$ are,
respectively, the scalar step function and a scalar ``top hat'' function. Thus, under gauge transformations,
the variation of the effective action can be written as
\be
-\delta W = \int dtdr \lambda \Big[\partial_r\Big(\frac{m^2}{4\pi}A_tH\Big) +\Big(J_{(o)}^r
-J_{(H)}^r +\frac{m^2}{4\pi}A_t\Big)\delta(r -r_H -\epsilon)\Big] \, ,
\ee
where we have omitted the contributions of the ingoing modes. If it is incorporated, the first
item is cancelled by its quantum effect at the horizon since its contribution to the total current is
$-m^2A_tH/(4\pi)$. In order to demand the full quantum theory to be gauge invariance at the quantum level,
the coefficient of the delta-function should vanish, that is
\be
a_o = a_H -\frac{m^2}{4\pi}A_t(r_H) \, ,
\ee
here we have applied the following expressions
\be
J_{(o)}^r = a_o \, , \qquad J_{(H)}^r = a_H +\frac{m^2}{4\pi}\big[A_t(r) -A_t(r_H)\big] \, ,
\ee
where $a_o$ and $a_H$ are the values of the consistent currents at infinity and the horizon, respectively.
To ensure the regularity requirement at the horizon, the covariant current should also vanish there. Since
a covariant current can be reformulated in the consistent form as $\widetilde J^r = J^r +\frac{m^2}{4\pi}
A_tH(r)$, the condition $\widetilde J^r = 0$ at the horizon determines the flux of the $U(1)$ gauge
current as
\be
a_o = -\frac{m^2}{2\pi}A_t(r_H) \, ,
\label{vc1}
\ee
which is required to cancel $U(1)$ gauge anomaly at the horizon and to restore the effective field theory
gauge invariance at the quantum level, and it is exactly equal to that of blackbody radiation at Hawking
temperature with an appropriate chemical potential. This is left to be discussed in the end of this section.

Next, we will study the flux of the energy momentum tensor. A gravitational anomaly is an anomaly in general
coordinate covariance, and explicitly expressed as the nonconservation of the energy momentum tensor. As
discussed above, the effective field theory is still defined in the region outside the event horizon. In
the region $r_H +\epsilon \leq r$, since there is an effective background gauge potential, but without any
anomaly, the energy momentum tensor satisfies the modified conservation equation: $\partial_rT_{(o)t}^r =
a_o\partial_r A_t$, which is deduced from the $\mu = t$ component of equation $\nabla_{\mu}T^{\mu}_t = 0$,
obeyed by the four-dimensional energy momentum tensor of the black hole. In the near horizon region $r_H
\leq r \leq r_H +\epsilon$, when omitting the classically irrelevant ingoing modes, the energy momentum
tensor exhibits an anomaly, and satisfies the anomalous equation as\footnote{In the two-dimensional reduction,
since its background has a gauge charge field, the energy momentum tensor is not conserved even at the
classical level. Under the general coordinate transformation $x\rightarrow x' = x -\xi$, the Ward identity
becomes $\nabla_\mu T_\nu^\mu = F_{\mu\nu}J^\mu +A_\nu\nabla_\mu J^\mu +\nabla_\mu N_\nu^\mu$, where we have
applied the transformations for the metric and the gauge field: $\delta g^{\mu\nu} = -(\nabla^\mu \xi^\nu
+\nabla^\nu \xi^\mu)$, $\delta A_\mu = \xi^\nu \partial_\nu A_\mu +\partial_\mu \xi^\nu A_\nu$.}
\be
\partial_rT_{(H)t}^r = J_{(H)}^r\partial_r A_t +A_t\partial_rJ_{(H)}^r +\partial_rN_t^r \, ,
\ee
where $N_t^r = (f_{,r}^2 +ff_{,rr})/(192\pi)$. As is shown above, for our discussions it is suffice to
consider the component $T_t^r$ for the energy momentum tensor only. The energy momentum tensor combines
contributions from these two regions, namely, $T_t^r = T_{(o)t}^r\Theta_+(r) +T_{(H)t}^rH(r)$. Under the
diffeomorphism transformation, the variation of the effective action reads off
\bea
-\delta W &=& \int dtdr \lambda^t\Big[a_o\partial_rA_t +\partial_r \Big(\frac{m^2}{4\pi}A_t^2
+N_t^r\Big)H(r) \nn \\
&& +\Big (T_{(o)t}^r -T_{(H)t}^r +\frac{m^2}{4\pi}A_t^2 +N_t^r\Big)\delta(r -r_H -\epsilon)\Big] \, ,
\eea
where we have used the relation $J_{(H)}^r = a_o +m^2A_t/(4\pi)$. The first term is the classical
effect of the background electric field for constant current flux. The second one is cancelled by
the quantum effect of the classically irrelevant ingoing modes whose contribution to the total energy
momentum is $-[m^2A_t^2/(4\pi) +N_t^r]H(r)$. The third one is eliminated by demanding the effective
action covariance under the diffeomorphism transformation, which permits
\be
c_o = c_H +\frac{m^2}{4\pi}A_t^2(r_H) -N_t^r(r_H) \, ,
\label{eq10}
\ee
in which $c_o = T_{(o)t}^r -a_oA_t$ and
\be
c_H = T_{(H)t}^r -\int_{r_H}^r dr \partial_r\big[a_oA_t +\frac{m^2A_t^2}{4\pi} +N_t^r\big]
\ee
are the values of the energy flow at the infinity and the horizon, respectively. Imposing a condition
\cite{MS,IUW2} that demands the covariant energy momentum tensor $\widetilde T_t^r = T_t^r +(ff_{,rr}
-2f_{,r}^2)/(192\pi)$ vanish at the horizon, we can obtain
\be
c_H = 2N_t^r(r_H) = \frac{\kappa^2}{24\pi} \, ,
\ee
where
\be
\kappa = \frac{1}{2}\partial_r f|_{r = r_H}
= \frac{r_H}{2(r_H^2+a^2)}\big(1 +\frac{3r_H^2 +a^2}{l^2} -\frac{a^2}{r_H^2}\big)
\ee
is the surface gravity of the black hole. The flux of the energy momentum tensor, which is required
to restore general coordinate covariance at the quantum level in the effective field theory, is
\be
c_o = \frac{\Xi^2 a^2 m^2}{4\pi(r_H^2+a^2)^2} +\frac{\pi}{12}T_H^2 \, ,
\label{vc2}
\ee
where $T_H = \kappa/(2\pi)$ is Hawking temperature of the black hole.

Now, we shall discuss Hawking fluxes of the black hole. To the simplicity, we focus on the fermions case, and then
 Hawking radiation of a Kerr-AdS black hole
is given by the Planckian distribution with a chemical potential, which is given by $N_{\pm m}(\omega)
= 1/[\exp(\frac{\omega\pm m A_t(r_H)}{T_H}) +1]$. With this distribution, the charge current and
energy momentum tensor fluxes of Hawking radiation are
\bea
F_m &=& m\int_0^\infty\frac{1}{2\pi}\Big[N_m(\omega) -N_{-m}(\omega)\Big]d\omega
= \frac{\Xi m^2 a}{2\pi(r_H^2+a^2)} \, , \label{tf1} \\
F_H &=& \int_0^\infty\frac{\omega}{2\pi}\Big[N_m(\omega) +N_{-m}(\omega)\Big]d\omega
= \frac{\Xi^2 a^2 m^2}{4\pi(r_H^2+a^2)^2} +\frac{\pi}{12}T_H^2 \, .
\label{tf2}
\eea
Comparing Eqs. (\ref{vc1}) and (\ref{vc2}), which are derived from the conditions of gauge and
gravitational anomaly cancellations and the regularity requirement at the horizon, with Eqs. (\ref{tf1})
and (\ref{tf2}), we find that the fluxes of Hawking radiation from a Kerr-AdS black hole are capable
of cancelling the anomalies at its horizon and restoring gauge invariance and general coordinate
covariance at the quantum level to hold in the effective theory. In fact, we can also obtain the same result in the dragging coordinate
system where the effective two-dimensional chiral theory contains only gravitational anomaly there,
resulting that the flux of Hawking radiation is only determined by the value of gravitational anomaly
at the horizon.

\section{Hawking radiation from a Kerr-Newman-AdS black hole}\label{KNAdS}

The line element of a Kerr-Newman-AdS black hole \cite{CP} takes the same form as that of a Kerr-AdS
black hole by replacing $2Mr$ with $2Mr -Q^2$ in the expreesion of $\Delta_r$. The black hole horizon
is determined by $\Delta_r(r_+) = 0$, and the potential of the electro-magnetic gauge field can be
written as
\be
\mathcal{A} = -\frac{Qr}{\rho^2}\big(dt -\frac{a\sin^2\theta}{\Xi} d\phi\big) \, .
\ee
By using a dimensional reduction similar to the previous section, the physics near the horizon can be
effectively described by an infinite collection of a $(1+1)$-dimensional complex scalar field in the
background with the same dilaton and metric as those presented in previous section, but now the $U(1)$
gauge field is given by
\be
A_t = -\frac{eQr}{r^2 +a^2} -\frac{\Xi ma}{r^2+a^2} \, .
\ee
The first term is originated from the electro-magnetic gauge potential of the charged black hole, while
the second one is derived from the induced gauge potential associated with the axisymmetry of the black
hole. So there are two gauge anomalies corresponding to, respectively, the original gauge symmetry and
the induced gauge symmetry when omitting the classically irrelevant ingoing modes at the horizon.
Correspondingly one need two gauge currents to cancel them. Following the procedure to determine the gauge charge
flux of a Kerr-AdS black hole, we first deal with the flux of the electric current. In the region $r_+
+\epsilon \leq r$, the electric current is conserved, and satisfies the conservation equation: $\partial_r
J_{(o)}^r = 0$. But in the region $r_+ \leq r \leq r_+ +\epsilon$, the effective two-dimensional chiral
theory contains the anomaly in gauge symmetry, and therefore the electric current satisfies the anomalous
equation, $\partial_r J_{(H)}^r = e\partial_r A_t/(4\pi)$. Under gauge transformations, the variation
of the effective action should vanish in order to restore gauge invariance at the quantum level. Finally,
the flux of the electric current, which is required to cancel the anomaly with respect to the original
gauge symmetry, can be determined by
\be
g_e = -\frac{e}{2\pi}{A_t}(r_+) = \frac{e^2Qr_+ +\Xi ema}{2\pi(r_+^2 +a^2)} \, .
\label{emf1}
\ee
In a similar way, one can obtain the $U(1)$ gauge charge flux, which is required to restore the
induced gauge symmetry at the quantum level, as
\be
g_m = -\frac{m}{2\pi}{A_t}(r_+) = \frac{meQr_+ +\Xi m^2a}{2\pi(r_+^2 +a^2)} \, .
\label{emf2}
\ee

Next, we will determine the flux of the energy momentum tensor. In the region $r_+ +\epsilon
\leq r$, the energy momentum tensor is conserved, and satisfies the conservation equation:
$\partial_r T_{(o)t}^r = \J_{(o)}^r \partial_r A_t$, which is related to the component $T_t^r$
of the four-dimensional energy momentum tensor in the black hole. In the near-horizon region
$r_+ \leq r \leq r_+ +\epsilon$, the energy momentum tensor exhibits an anomaly, and satisfies
the anomalous equation, $\partial_r T_{(H)t}^r = \J_{(H)}^r \partial_r A_t +A_t\partial_r
J_{(H)}^r +\partial_r N_t^r$. In both regions, we define the current $\J^r \equiv J^r/e = j^r/m$,
where the current $j^r$ is originated from the axial symmetry and deduced from the component
$T_\phi^r$ of the four-dimensional energy momentum tensor in the black hole. Adopting the same
procedure as previous section, the flux of the energy momentum tensor is given by
\be
g_o = \frac{1}{4\pi}A_t^2(r_+) +N_t^r(r_+)  =\frac{1}{4\pi} \Big(\frac{eQr_+}{r_+^2 +a^2}
+\frac{\Xi ma}{r_+^2 +a^2}\Big)^2 +\frac{\pi}{12}T_+^2 \, ,
\label{emf3}
\ee
where
\be
T_+ = \frac{r_+}{4\pi(r_+^2+a^2)}\big(1 +\frac{3r_+^2 +a^2}{l^2} -\frac{a^2 +Q^2}{r_+^2}\big)
\ee
is Hawking temperature of the black hole. The Hawking radiant spectrum of the Kerr-Newman-AdS
black hole is given by the Planckian distribution: $N_{\pm e, \pm m}(\omega) = 1/[\exp(
\frac{\omega \mp e\Psi_+ \mp m\Omega_+}{T_+}) +1]$ for the fermion case, where $\Psi_+ = Qr_+/(r_+^2 +a^2)$ and
$\Omega_+ = \Xi a/(r_+^2 +a^2)$. From this distribution, the electric current flux, the angular
momentum flux(namely, the $U(1)$ gauge current flux) and the energy momentum tensor flux of
Hawking radiation can be obtained as
\bea
F_e &=& e\int_o^\infty \frac{1}{2\pi}[N_{e, m}(\omega) -N_{-e, -m}(\omega)]d\omega
= \frac{e^2Qr_+ +\Xi ema}{2\pi(r_+^2 +a^2)} \, , \label{fhr1} \\
F_m &=& m\int_o^\infty \frac{1}{2\pi}[N_{e, m}(\omega) -N_{-e, -m}(\omega)]d\omega
= \frac{meQr_+ +\Xi m^2a}{2\pi(r_+^2 +a^2)} \, , \label{fhr2} \\
F_H &=& \int_o^\infty \frac{\omega}{2\pi}[N_{e, m}(\omega) +N_{-e, -m}(\omega)]d\omega
= \frac{1}{4\pi}\Big(\frac{eQr_+}{r_+^2 +a^2} +\frac{\Xi ma}{r_+^2 +a^2}\Big)^2
+\frac{\pi}{12}T_+^2 \, . \nn \\
&& \label{fhr3}
\eea

Comparing Eqs. (\ref{emf1}-\ref{emf3}) with Eqs. (\ref{fhr1}-\ref{fhr3}), we conclude that the fluxes
of the electric current, the $U(1)$ gauge current and the energy momentum tensor, required to cancel
gauge or gravitational anomalies at the horizon, are equal to that of Hawking radiation.

In the following, we shall investigate Hawking radiation via anomalous point of view in the dragging
coordinate system. As is shown above, by using a dimensional reduction technique to a four-dimensional
Kerr-Newman-AdS black hole in the Boyer-Lindquist coordinates $(t, \phi)$, the two-dimensional effective theory for each
partial wave has an original gauge symmetry with respect to the electric field of the black hole and an
induced $U(1)$ gauge symmetry originated from the isometry along the $\phi$-direction. The chiral theory
then contains two gauge anomalies and a gravitational anomaly at the horizon when omitting classically the
ingoing modes. The flux of Hawking radiation can be treated by the cancellation conditions of these anomalies
at the horizon of the black hole. If instead, we adopt the dragging coordinates $(\xi, \psi)$ where the
matter field in the ergosphere at the horizon must be dragged by the gravitational field with an azimuthal
angular momentum, rather than use the coordinates $(t, \phi)$, the effective near-horizon quantum field
contains only one gauge anomaly associated with the electric field of the black hole, and a gravitational
anomaly after excluding mode's propagation along one light-like direction. Specifically, after undergoing
a dimensional reduction technique to the rotating charged black hole in the dragging coordinates, physics
near the horizon can be effectively described by an infinite collection of $(1+1)$-dimensional field in
the background of the dilaton $\Psi$, the metric and the gauge field $A_{\mu}$
\bea
ds^2 &=& -f(r)d\xi^2 +\frac{1}{f(r)}dr^2 \, , \qquad f(r) = \frac{\Delta_r}{r^2+a^2} \, , \nn \\
\Psi &=& \frac{r^2 +a^2}{\Xi} \, , \qquad A_t = -\frac{Qr}{r^2 +a^2} \, , \quad A_r = 0 \, ,
\eea
where we have introduced the dragging coordinate $(\xi, \psi)$ through $\psi = \phi -\Omega_+t$, $\xi = t$,
and $e$ is the charge of the test field. Following the procedure done in the previous section to calculate
the value of the charge current flux, the effective theory is also defined outside the horizon. In the region
$r_+ +\epsilon \leq r$, the electric current is conserved, and satisfies the conservation equation $\partial_r
J_{(o)}^r = 0$. Near the horizon, when omitting the classically irrelevant ingoing modes, the effective quantum
field theory becomes chiral, the electric current exhibits an anomaly, and satisfies equation $\partial_rJ_{(H)}^r
= e^2\partial_r A_t/(4\pi)$. Under gauge transformations, physical condition and the effective action should
be gauge invariant at the quantum level, both the covariant form of the electric current and the variation
of the effective action should vanish there. Thus the flux of the electric current is given by
\be
H_o = -\frac{e^2}{2\pi}A_t(r_+) = \frac{e^2Qr_+}{2\pi(r_+^2 +a^2)} \, .
\ee

Next, we will study the flux of the energy momentum tensor. In the region $r_+ +\epsilon \leq r$, the energy
momentum tensor is conserved, and satisfies equation $\partial_r T_{(o)t}^r = J_{(o)}^r\partial_rA_t$, while
in the region $(r_+ \leq r \leq r_+ +\epsilon)$, the effective theory is chiral, the energy momentum tensor
exhibits an anomaly and satisfies the anomalous equation
\be
\partial_r T_{(H)t}^r = J_{(H)}^r\partial_rA_t +A_t\partial_r J_{(H)}^r +\partial_rN_t^r \, .
\ee
In order to restore the covariance of the effective action under diffeomorphism transformations, and the
regularity of physical condition at the horizon, the covariant the energy momentum tensor should be
conserved and the variation of the effective action should vanish in the near-horizon region. Thus one
can obtain
\be
K_o = \frac{e^2}{4\pi}A_t^2(r_+) +N_t^r(r_+)
= \frac{e^2Q^2r_+^2}{4\pi{(r_+^2 +a^2)}^2} +\frac{\pi}{12}T_+^2 \, ,
\ee
where $K_o$ is the flux of the energy momentum tensor, and $T_+ = \kappa/(2\pi)$ is Hawking temperature at
the outer horizon. In the dragging coordinates, the Hawking distribution is given by Planckian distribution
with an electric chemical potential, namely, $N_{\pm e}(\varpi) = 1/[\exp({\frac{\varpi\pm eA_t(r_+)}{T_+}})
+1]$ for the fermion case, where $\varpi=\omega-m\Omega_+$ is the energy carried by the observer in the dragging
coordinate system. So the electric current and energy momentum tensor fluxes of Hawking radiation are given by
\bea
F_e &=& e\int_o^\infty \frac{1}{2\pi}[N_e(\varpi) -N_{-e}(\varpi)]d\varpi
= \frac{e^2Qr_+}{2\pi(r_+^2 +a^2)} \, , \label{eq23} \nn \\
F_H &=& \int_o^\infty \frac{\varpi}{2\pi}[N_{e}(\varpi) +N_{-e}(\varpi)]d\varpi
= \frac{e^2Q^2r_+^2}{4\pi{(r_+^2 +a^2)}^2} +\frac{\pi}{12}T_+^2 \, .
\eea

Obviously, in the coordinates $(\xi, \psi)$, the electric current and energy momentum tensor fluxes of Hawking
radiation have equivalent forms to those obtained by requiring gauge invariance and general coordinate
covariance at the quantum level to hold in the effective theory. However, an observer rest at the dragging coordinate system would not
observe the $U(1)$ gauge current flux that originates from the axisymmetry of the black hole since the
observer in the dragging coordinate system is co-rotating with the axisymmetric black hole, and resulting
that the effective theory does not contain the $U(1)$ gauge anomaly at the horizon. If we restore the Boyer-Lindquist
coordinate system from the dragging coordinate system, it is also easy to calculate the flux of the
angular momentum\cite{MS}. This result supports the Robinson-Wilczek's opinion, and further extends
it to the rotating black hole in the dragging coordinate system.

\section{Discussions and Conclusions}\label{dc}

An anomaly is originated from a conflict between a symmetry of classical action or the corresponding conservation
law and the procedure of quantization. Specifically, under gauge or diffeomorphism transformations, the symmetry
of the classical action is expressed as the covariant conservation of the corresponding gauge current or energy
momentum tensor, but when moving to the quantum theory, the gauge current or energy momentum tensor is no longer
conserved due to the quantum effect taken place at the horizon, and displays the chiral anomaly on the horizon.
To relieve the conflict, gauge or gravitational anomaly arises from the fact that one wants to exclude those
modes propagating along one light-like direction.

In the case of a rotating Kerr black hole in the AdS spaces, by using a dimensional reduction technique, the
effective two-dimensional theory for each partial wave has a $U(1)$ gauge symmetry corresponding to the isometry
along $\phi$-direction and quantum field in the original spacetime can now be interpreted as a $(1+1)$-dimensional
charged field with a charge proportional to the azimuthal angular momentum $m$. When we omit the classically
irrelevant ingoing modes, $U(1)$ gauge anomaly with respect to the axisymmetry of the black hole arises at the
horizon. In the case of a Kerr-Newman-AdS black hole, there are two gauge symmetries for each partial wave, one
is the primary gauge symmetry originated from the electric field radiated from the charged black hole, the other
is the induced symmetry which is associated with the isometry along the $\phi$-direction. Correspondingly, there
are two gauge currents to cancel these anomalies at the horizon when omitting the classically irrelevant
ingoing modes. However, the induced gauge symmetry in the field radiated from the axisymmetry black hole
is masked in the dragging coordinate system since the dragging coordinate system corresponds to a kind of
locally non-rotating coordinate system in which an observer lives would not observe this $U(1)$ gauge flux
radiated from the horizon.

In summary, we have investigated Hawking radiation of the four-dimensional Kerr-AdS and Kerr-Newman-AdS black
holes from the viewpoint of cancelling gravitational or gauge anomalies at the horizon. The results show
that the fluxes of gauge current or energy momentum tensor, which are required to cancel gauge or
gravitational anomalies at the horizon, are exactly equal to that of blackbody radiation at
Hawking temperature with appropriate chemical potentials and the thermal flux of Hawking radiation is capable
of cancelling the anomaly at the horizon. This generalizes and supports Robinson-Wilczek' method to derive
Hawking radiation via anomalies point of view.

\section*{Acknowledgments}
This work was partially supported by the Natural Science Foundation of China under Grant No. 10675051,
10635020, 70571027, 70401020 and a grant by M.O.C under Grant No. 306022. S.Q.-Wu was also supported
in part by a starting fund from Central China Normal University.

\end{document}